\documentclass[
 aip,
 pop,
amsfonts,
amsmath,
amssymb,
reprint,%
]{revtex4-1}

\usepackage{textcomp}
\usepackage{amsmath,amstext,amsfonts,amssymb, bm}
\usepackage[colorlinks,linkcolor=blue,urlcolor=blue,citecolor=blue]{hyperref}
\usepackage{graphicx}
\usepackage{gensymb}

\usepackage{subfig}

\usepackage{xcolor}
\usepackage{lineno}
\usepackage{multirow}
\usepackage{dcolumn}
\usepackage{etoolbox}

\makeatletter
\def\@email#1#2{%
 \endgroup
 \patchcmd{\titleblock@produce}
  {\frontmatter@RRAPformat}
  {\frontmatter@RRAPformat{\produce@RRAP{*#1\href{mailto:#2}{#2}}}\frontmatter@RRAPformat}
  {}{}
}%
\makeatother
\begin{document}

\title{Shock Hugoniot calculations using on-the-fly machine learned force fields with ab initio accuracy}

\author{Shashikant Kumar}
 \affiliation{College of Engineering, Georgia Institute of Technology, Atlanta, GA 30332, USA}
 \author{John E. Pask}
\affiliation{Physics Division, Lawrence Livermore National Laboratory, Livermore, CA 94550, USA}
\author{Phanish Suryanarayana}
\email[Email: ]{phanish.suryanarayana@ce.gatech.edu}
\affiliation{College of Engineering, Georgia Institute of Technology, Atlanta, GA 30332, USA}
\affiliation{College of Computing, Georgia Institute of Technology, Atlanta, Georgia 30332, USA}

\date{\today}

\begin{abstract}
We present a framework for computing the shock Hugoniot using on-the-fly machine learned force field (MLFF) molecular dynamics simulations. In particular, we employ an MLFF model based on the kernel method and Bayesian linear regression to compute the electronic free energy, atomic forces, and pressure; in conjunction with a linear regression model between the electronic internal and free energies to compute the internal energy, with all training data generated from Kohn-Sham density functional theory (DFT). We verify the accuracy of the formalism by comparing the Hugoniot for carbon with recent Kohn–Sham DFT results in the literature. In so doing, we demonstrate that Kohn-Sham calculations for the Hugoniot can be accelerated by up to two orders of magnitude, while retaining \emph{ab initio} accuracy. We apply this framework to calculate the Hugoniots of 14 materials in the FPEOS database, comprising 9 single elements and 5 compounds, between temperatures of 10 kK and 2 MK. We find good agreement with first principles results in the literature while providing tighter error bars. In addition, we confirm that the inter-element interaction in compounds decreases with  temperature. 
\end{abstract}


\maketitle

\section{\label{sec:introduction}Introduction}

Dynamic shock compression experiments \cite{eggert2008Hugoniot, knudson2003use, fortov2007phase} are commonly used for investigating material properties and behavior under extreme conditions of temperature and pressure, such as those encountered in the warm dense matter (WDM) and hot dense matter (HDM) regimes.  The shock Hugoniot,  the locus of all possible final shocked states for a given initial state,  has a number of applications. These include  understanding the dynamic behavior and shock-induced reactions during planetary impacts \cite{umeda2022hugoniot}, modeling of shock initiation and detonation in polymer-bonded explosives \cite{zhou2024shock}, and investigation of implosions for inertial confinement fusion \cite{gaffney2018review}.  However, given the significant cost and complexity of such experiments, Hugoniot data for different materials is limited.

Kohn-Sham density functional theory (DFT) \cite{kohn1965self, hohenberg1964inhomogeneous} is among the most widely used electronic structure methods for understanding and predicting the properties of materials from the first principles of quantum mechanics, with no empirical parameters. However, Kohn-Sham DFT calculations are associated with significant computational costs, particularly for  ab initio molecular dynamics (AIMD) simulations, where it is common to solve the Kohn-Sham equations many thousands of times and more. There is also a rapid increase in  cost with temperature as the number of partially occupied states increases, which has limited the application of Kohn-Sham DFT for Hugoniot calculations to a relatively small number of cases \cite{hu2016firstP, bernard2002first, driver2015first,driver2015firstNeon, joshi2009shock, zhao2015first, zhang2017first, strickson2016ab, wixom2011calculating, zhang2017equation, pu2021molecular, romero2007density, hu2014properties, militzer2006first, yang2018molecular, driver2015firstP, root2010shock}, with a significant number of them restricted to a small temperature range at the lower end of the spectrum \cite{bernard2002first, joshi2009shock, zhao2015first, pu2021molecular, strickson2016ab}.   Though  path integral Monte Carlo (PIMC) \cite{barker1979quantum} is an alternative first principles method that can be used for the calculation of the Hugoniot \cite{driver2012all, militzer2000pathPIMC, driver2018path, militzer2015development}, it is associated with significant computational costs that rapidly increase as the  temperature decreases, making it impractical at the lower temperatures, particularly for systems with heavier chemical elements. 

There have been recent efforts directed at developing solution strategies for Kohn-Sham DFT that are tailored to calculations at high temperature \cite{suryanarayana2018sqdft, pratapa2016spectral, cytter2018stochastic, baer2013self, white2020fast, xu2022real, zhang2016extended, blanchet2021extended, sadigh2023spectral}. Though they represent significant advances, the methods still entail considerable computational cost, particularly for the AIMD simulations required in shock Hugoniot calculations. This limitation can be overcome using machine learned force field (MLFF) schemes \cite{unke2021machine, poltavsky2021machine, wu2023applications}. In particular, in their original form,   MLFFs have found application for temperatures up to the lower part of the WDM regime \cite{liu2020structure, mahmoud2022predicting, kumar2023transferable, zeng2021ab, shi2023double, willman2020quantum, richard2023ab, willman2021carbon}. These schemes have been extended to include the prediction  of the internal energy \cite{zhang2020warm, tanaka2022development, ben2022predicting, fiedler2023machine, ellis2021accelerating, ben2020learning}, making them amenable to the study of the WDM and HDM regimes, including the calculation of the shock Hugoniot \cite{zhang2020warm}.  However, these methods require extensive training datasets comprised of DFT data from thousands of calculations. Such a process is not only computationally and labor-intensive,  but also needs to be repeated for different conditions. This limitation can be overcome using on-the-fly MLFF, where training of the model  is performed during the molecular dynamics (MD) simulation itself \cite{jinnouchi2019fly, jinnouchi2020fly, verdi2021thermal, liu2021alpha, Chen2023transferable, kumar2023kohn, timmerman2024overcoming}. Such a scheme has been applied to calculate the transport properties and WDM and HDM \cite{kumar2024fly, stanek2024review}.  However, it does not provide the internal energy, as required for shock Hugoniot calculations.

In this work, we present a framework for calculating the shock Hugoniot using on-the-fly MLFF MD simulations. In particular, we employ an MLFF model based on the kernel method and Bayesian linear regression to compute the electronic free energy, atomic forces, and pressure; along with a model based on linear regression  between the electronic internal and free energies to compute the internal energy, with all training data generated from Kohn-Sham DFT. Using this framework, we demonstrate that shock Hugoniot calculations can be accelerated by two orders of magnitude while retaining Kohn-Sham accuracy. We apply this framework to the calculation of the shock Hugoniots for 14 materials, where we find good agreement with the FPEOS database \cite{militzer2021firstprin}. In so doing, we confirm that the inter-element interaction in compounds decreases with  temperature.

The remainder of this paper is organized as follows. In Section~\ref{sec:formulation}, we discuss the formulation and implementation of the on-the-fly MLFF scheme for shock Hugoniot calculations. In Section~\ref{Sec:Results}, we verify the accuracy and performance of the framework and then apply it  to compute the shock Hugoniots for various materials.  Finally, we provide concluding remarks in Section~\ref{sec:conclusion}.


\section{\label{sec:formulation}Formulation and Implementation}

The  Rankine-Hugoniot relation for the states of matter on either side of a shock front can be written as: 
    \begin{equation} 
       u_1- u_0 = \frac{1}{2}(P_1+P_0)\left(\frac{1}{\rho_0} - \frac{1}{\rho_1}\right) \,,
        \label{Eq:Hugoniot}
    \end{equation}
where $u$ is the internal energy, $P$ is the pressure, and $\rho$ is the density, with the subscripts 0 and 1 denoting the initial and shocked states, respectively. Given an initial state, typically corresponding to ambient conditions,  Eq.~\ref{Eq:Hugoniot} can be solved along different isotherms to determine the Hugoniot, i.e., for a given temperature $T$, the values of $\rho_1$ and the corresponding $P_1$ within the material's equation of state (EOS) are determined such that the Rankine-Hugoniot condition  is satisfied. Therefore, the construction of the Hugoniot involves EOS calculations at a number of shocked states. 

\begin{figure}[htbp]
\centering
\includegraphics[width=0.98\linewidth]{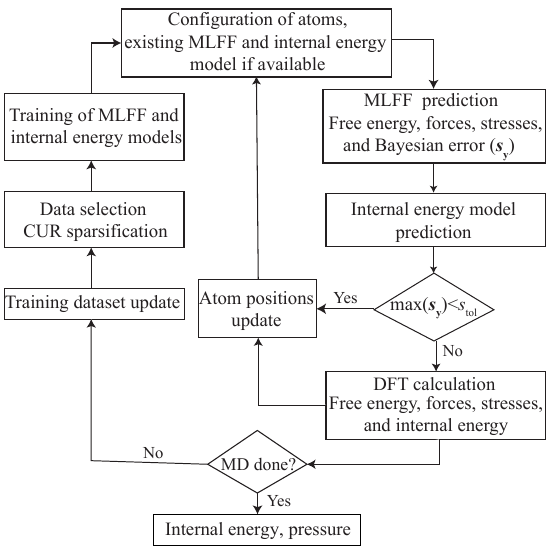}
        \caption{Outline of the on-the-fly MLFF MD framework for shock Hugoniot calculations.}
        \label{fig:flowchart}
\end{figure}

In this work, extending recent developments \cite{kumar2023kohn, kumar2024fly}, the EOS is calculated using the on-the-fly MLFF MD framework outlined in Fig.~\ref{fig:flowchart}, which is   implemented in parallel within the SPARC electronic structure code \cite{xu2021sparc, zhang2023version}, building on a  prototype version in its serial MATLAB counterpart, M-SPARC \cite{xu2020m, zhang2023versionM}. The MD simulation starts with a series of Kohn-Sham DFT calculations that provide the initial training data for the free energy, atomic forces, and stresses. The machine learned models are then used to predict the internal energy, forces, and stresses in subsequent MD steps,  except when the Bayesian uncertainty/error in the forces so computed exceeds the threshold $s_{tol}$, at which point a DFT calculation is performed and the quantities so calculated are added to the training dataset. The threshold $s_{tol}$ is dynamic, set to the maximum value of the Bayesian error in forces for the MLFF  step subsequent to the DFT training step. To mitigate the cubic scaling bottleneck encountered during training, a two-step data selection method is employed: atoms with Bayesian force errors exceeding a set threshold are included in the training dataset, followed by CUR \cite{mahoney2009cur} for downsampling.

The Kohn-Sham training data is generated using the SPARC code, where we employ diagonalization based algorithms for simulations at temperatures less than 100 kK and employ the Spectral Quadrature (SQ) method \cite{bhattacharya2022accurate, suryanarayana2018sqdft, pratapa2016spectral, suryanarayana2013coarse} for higher temperatures. Diagonalization based algorithms solve the Kohn-Sham equations for the occupied orbitals, which scales cubically with the number of orbitals. Since the number of orbitals that need to be calculated increases with temperature, as determined by the Fermi-Dirac distribution, the cost of such a procedure increases rapidly with temperature. In the Gauss SQ method, the electronic density and free energy are written as bilinear forms or sums of bilinear forms, which are then approximated by Gauss quadrature rules. In so doing, the  Hellmann-Feynman atomic forces and stresses  also become available \cite{kumar2024fly, sharma2020real, pratapa2016spectral, suryanarayana2018sqdft}. On exploiting the locality of electronic interactions \cite{prodan2005nearsightedness}, which manifests itself as the exponential decay of the finite-temperature density matrix away from its diagonal \cite{goedecker1998decay, ismail1999locality, benzi2013decay, suryanarayana2017nearsightedness}, the SQ method scales linearly with system size. The associated prefactor decreases rapidly with temperature, making it the preferred method for Kohn-Sham calculations at high temperature \cite{kumar2024fly, bethkenhagen2023properties, wu2021development, zhang2019equation}. Note that in addition to the standard parameters in DFT calculations, there are two additional parameters in the SQ method, namely the truncation radius and the quadrature order, both of which can be systematically converged to the accuracy desired.

In the MLFF model, the electronic free energy admits the following decomposition \cite{bartok2010gaussian, kumar2024fly}:
\begin{equation} 
    \mathcal{F}^{\rm MLFF} = \sum_{E} \sum_{i=1}^{N_A^E} \varphi_i^E  = \sum_E \sum_{i=1}^{N_A^E} \sum_{t=1}^{N_{T}^{E}} \tilde{w}_{t}^{E} k\left(\mathbf{x}_{i}^{E}, \tilde{\mathbf{x}}_t^{E} \right) \,,
    \label{Eq:EnergyDecompML}
\end{equation}
where $\varphi_i^E$ are the atomic free energies, $\tilde{w}_{t}^{E}$ are the model weights, and $k\left(\mathbf{x}_{i}^{E}, \tilde{\mathbf{x}}_t^{E} \right)$ is a kernel function that assesses the similarity between descriptor vectors $\mathbf{x}_{i}^{E}$ and $\tilde{\mathbf{x}}_t^{E}$, with $N_A^E$ denoting the number of atoms in the current atomic configuration, $N_{T}^{E}$ representing the number of atoms in the training dataset, and $E$ representing the chemical element, which is also used as an index.  In this work, we employ the polynomial kernel in conjunction with  Smooth Overlap of Atomic Positions (SOAP)  descriptors \cite{bartok2013representing}. Since the descriptors have explicit dependence on the atomic positions, the atomic forces and stresses corresponding to the free  energy model in Eq.~\ref{Eq:EnergyDecompML} are readily obtained \cite{kumar2023kohn}. During training, the weights are computed using Bayesian linear regression \cite{bishop2006pattern}:
\begin{equation} 
    \tilde{\boldsymbol{w}} = \beta(\alpha \boldsymbol{I} + \beta \tilde{\boldsymbol{K}}^{T} \tilde{\boldsymbol{K}})^{-1}  \tilde{\boldsymbol{K}}^{T} \tilde{\boldsymbol{y}} \,,
    \label{Eq:wtsML}
\end{equation}
where $\tilde{\boldsymbol{w}}$ is the vector of weights;  the parameters $\alpha$ and $\beta$ are chosen to maximize the evidence function \cite{bishop2006pattern}; $\tilde{\boldsymbol{y}}$ is a vector of the energy, atomic forces, and stresses for the atomic configurations in the training dataset, after having been shifted and normalized by their mean and standard deviation values, respectively \cite{kumar2023kohn};  and $\tilde{\boldsymbol{K}}$ is the corresponding covariance matrix \cite{kumar2023kohn}. After training, the energy, atomic forces, and stresses can be predicted as: 
\begin{equation} 
    \boldsymbol{y} = \boldsymbol{K} \tilde{\boldsymbol{w}} \,,
    \label{Eq:LinearSystemML}
\end{equation}
where $\boldsymbol{K}$ represents the covariance matrix associated with the given atomic configuration.  The uncertainty in the predicted values can be calculated using the relation:
\begin{equation} 
    \boldsymbol{s}_{\boldsymbol{y}} = \sqrt{{\rm diag}\left( \frac{1}{\beta} \boldsymbol{I} + \boldsymbol{K} (\alpha \boldsymbol{I} + \beta \tilde{\boldsymbol{K}}^{T} \tilde{\boldsymbol{K}})^{-1} \boldsymbol{K}^{T} \right)}  \,,
    \label{Eq:UQML}
\end{equation}
where $\mathbf{s}_{\mathbf{y}}$ is a vector with entries corresponding to the Bayesian errors in the energy, atomic forces, and stresses. 

The MLFF model described above only provides access to the electronic free energy and not the internal energy, as required for the calculation of the shock Hugoniot (Eq.~\ref{Eq:Hugoniot}). In this work, we employ an internal energy model based on linear regression between the free and internal energies. We find that higher-order regression provides negligible improvement in the model, and so linear regression suffices. As in the MLFF model, this internal energy model is trained on the steps at which a Kohn-Sham DFT calculation is performed.  An alternative  to this strategy is  to use a model for the internal energy  that is similar to the free energy (Eq.~\ref{Eq:EnergyDecompML}).  However, since each atomic configuration provides a single value for the internal energy, whose derivatives with respect to atom positions are not available, a relatively large number of DFT steps will be required to train the model, which will  significantly increase the cost of the MD simulation.  The strategy adopted here provides the desired accuracy, as demonstrated in the next section, with negligible increase in the cost of the MD simulation.


\section{\label{Sec:Results} Results and discussion}
We first verify the accuracy and performance of the developed MLFF formalism in Section~\ref{sec:accuracy_performace}. Next, we apply the framework to calculate the Hugoniots for various materials  in Section~\ref{sec:FPEOScompare}. The data for the Hugoniots so calculated can be found in Appendix~\ref{App:Data}, while the data for all the  EOS calculations used in forming each  Hugoniot can be found in the Supplementary material. In calculating the Hugoniot,  the pressure is calculated for 4 densities at each of the temperatures chosen, and then the point on the Hugoniot is determined by performing linear interpolation between the density and the residual of the Rankine-Hugoniot relation (Eq.~\ref{Eq:Hugoniot}). Given the proximity of the chosen densities to the corresponding one on the Hugoniot, the quality of the fit is excellent, with $R^2 \geq 0.99$ being the coefficient of determination for the linear regression in all cases.

We compute the pressure by performing isokinetic ensemble (NVK) MD simulations with a Gaussian thermostat \cite{minary2003algorithms} for 10,000 steps, with the initial 1,000 steps used for equilibration and the remaining 9,000 used for calculation of the pressure. In so doing, the statistical error in the pressure is reduced to less than 0.1\%. The starting atomic configurations for each MD simulation are generated through a preliminary equilibration step in which an on-the-fly MLFF MD simulation of 1000 steps is performed. For the Kohn-Sham DFT calculations performed during this step, we not only choose the grid spacings to be twice larger than those used for training, but also choose other numerical parameters to be significantly less strict, whereby the associated computational cost  is negligible compared to the cost of the production MD simulation. This procedure ensures that the machine learned models are trained only after some initial equilibration, which helps to increase the accuracy of the resulting model.

In the Kohn–Sham calculations performed during the on-the-fly MLFF scheme, we use the Perdew–Burke–Ernzerhof (PBE) \cite{perdew1996generalized} exchange-correlation functional, unless  otherwise specified. In addition, we use optimized norm-conserving Vanderbilt (ONCV) pseudopotentials \cite{hamann2013optimized}, chosen from the SPMS set \cite{shojaei2023soft} for temperatures lower than 100,000 K, and from previous work \cite{bethkenhagen2023properties, suryanarayana2023accuracy, kumar2023kohn} for higher temperatures. The internal energy and pressure for the initial state, i.e., $u_0$ and $P_0$, respectively, are calculated using their respective  pseudopotentials to ensure that the initial and shocked states have a common reference while  using Eq.~\ref{Eq:Hugoniot}.  All the numerical parameters in DFT, including the grid spacing in the standard diagonalization calculations, and the grid spacing, truncation radius, and quadrature order in the SQ calculations,  are chosen such that the pressures are converged to within 1\%. Indeed, the errors can be reduced as desired by choosing smaller grid spacings, and in the case of SQ, also larger truncation radii and quadrature orders. In the MLFF calculations, the hyperparameters are chosen to be the same as in previous work \cite{kumar2024fly, kumar2023kohn}. Indeed, we have found that the accuracy and performance of the MLFF scheme is insensitive to the choice of these parameters for the range of materials systems and conditions studied here. 

\subsection{\label{sec:accuracy_performace}Accuracy and performance}

We now compute the shock Hugoniot for carbon between temperatures of 10 kK and 10 MK, containing both the WDM and HDM regimes. In particular, we consider the following temperatures: 10 kK, 20 kK, 50 kK, 100 kK, 200 kK, 500 kK, 750 kK, 1 MK, 2 MK, 5 MK, and 10 MK. The  temperature and density for the initial state are chosen to be 300 K and 3.515 g/cm$^3$, respectively. In the MD simulations, we employ a time step of 0.63 fs for the temperature of 10 kK, adjusting it for other temperatures as the inverse square root of the temperature. We consider system sizes of 500 and 64 atoms for the lowest and highest temperatures of 10 kK and 10 MK, respectively, while interpolating linearly for temperatures in between. We employ the local density approximation (LDA) \cite{kohn1965self} for the exchange-correlation functional to facilitate comparison with previous DFT results. 

\begin{figure}[htbp]
\includegraphics[width=0.85\linewidth]{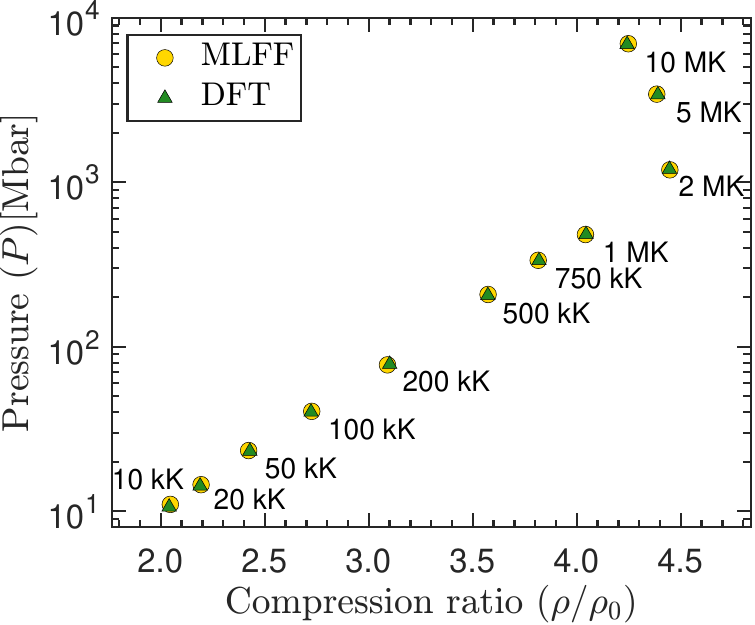}
        \caption{Shock Hugoniot for carbon obtained from the on-the-fly MLFF scheme and Kohn-Sham DFT \cite{bethkenhagen2023properties}.}
        \label{fig:validation}
\end{figure}

\begin{figure*}[htbp]
\subfloat[$\rho$ = 8.5 g/cm$^3$]{\includegraphics[keepaspectratio=true,width=0.24\textwidth]{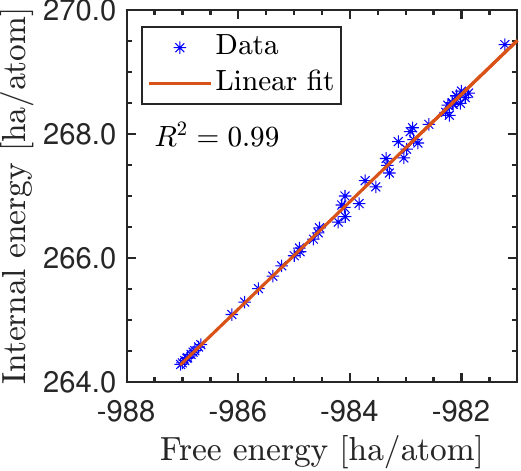} \label{Fig:regression_entropy_1} }
\subfloat[$\rho$ = 10 g/cm$^3$]{\includegraphics[keepaspectratio=true,width=0.24\textwidth]{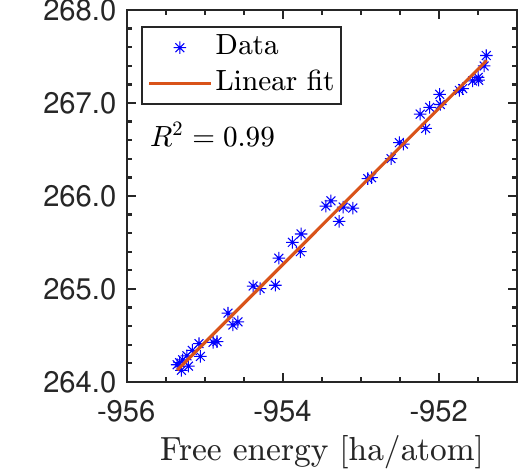} \label{Fig:regression_entropy_2} }
\subfloat[$\rho$ = 13 g/cm$^3$]{\includegraphics[keepaspectratio=true,width=0.24\textwidth]{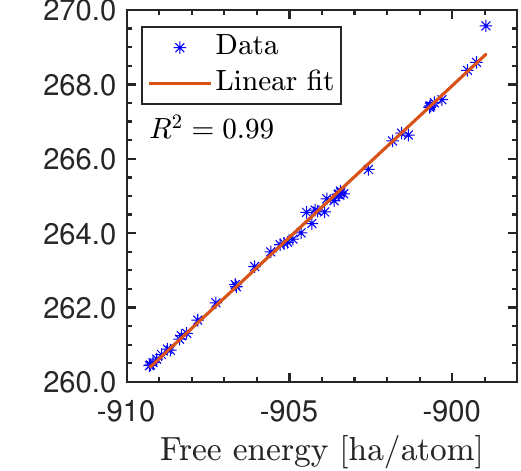} \label{Fig:regression_entropy_3} }
\subfloat[$\rho$ = 16 g/cm$^3$]{\includegraphics[keepaspectratio=true,width=0.24\textwidth]{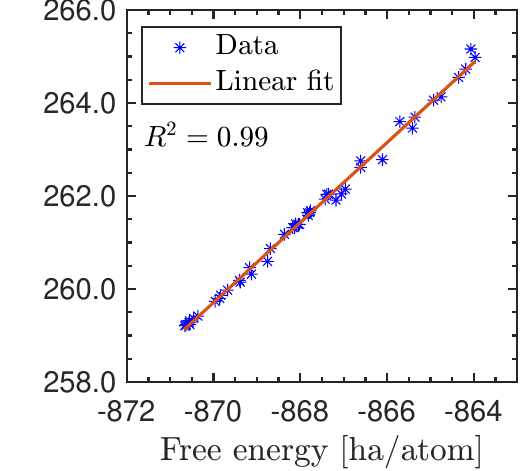} \label{Fig:regression_entropy_4} }
\caption{\label{Fig:regression_entropy} Internal energy model for carbon at a temperature of 10 MK and different densities.  The $R^2$ value represents the coefficient of determination for the linear regression.}
\end{figure*}

In Fig.~\ref{fig:validation}, we present the shock Hugoniot of carbon so computed using the on-the-fly MLFF scheme, and comparison to recent Kohn-Sham DFT results in the literature \cite{bethkenhagen2023properties}. We observe that there is very good agreement between the MLFF scheme and  Kohn-Sham DFT, with a maximum difference  of 1.9\% in the pressure and 0.4\% in the density along the Hugoniot. The corresponding errors in the pressure and internal energy predicted by the MLFF scheme, as determined by averaging the error over all the DFT steps,  are 1.4\% and 0.005 ha/atom, respectively. In particular, the error in the internal energy is dominated by the error in the MLFF model for the free energy, given the excellent   fit in the internal energy model, as shown in Fig.~\ref{Fig:regression_entropy}. These results indicate that the MLFF scheme can be used to reliably  calculate the shock Hugoniot over a wide range of temperatures. Note that it is possible to formulate the MLFF model directly in terms of the pressure, rather than the different components of the stress tensor, as done here. We have found both strategies to provide similar accuracies, e.g., for a temperature of 500 kK, the formalisms in terms of the pressure and stress tensor result in errors of   1.1\% and 0.8\%, respectively, in the pressure. Note also that the internal energy model can be trained subsequent to the complete MD simulation, i.e., as a post-processing step. We have found that this strategy provides similar accuracy to that adopted here. Indeed, the current strategy of on-the-fly training of the internal energy model accounts for changes in the response of the system over the course of the MD simulation. 

\begin{figure}[htbp]
\includegraphics[width=0.75\linewidth]{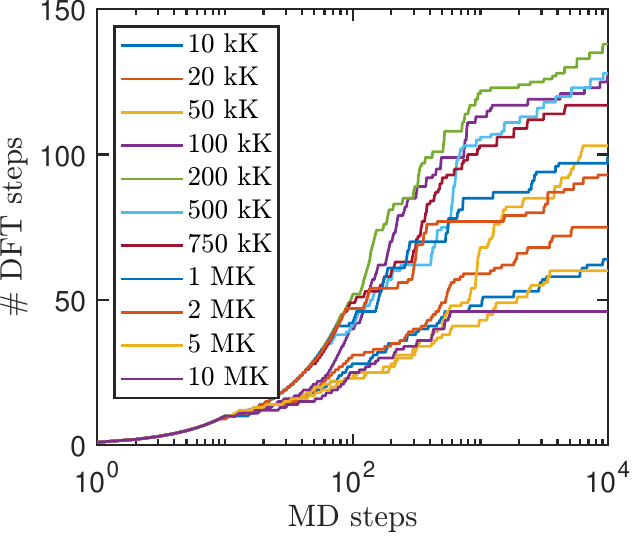}
        \caption{Cumulative number of DFT steps during the on-the-fly MLFF MD simulation for carbon at different temperatures.}
        \label{fig:DftCumSteps}
\end{figure}

In Table~\ref{Tab:MLFFPerformanceTable}, we present the performance of the on-the-fly MLFF scheme in performing the  MD simulations above. We observe that the number of DFT steps constitutes  less than 1.4\% of the total number of MD steps, with the minimum and maximum percentages being 0.46\% and 1.38\%, respectively.  The occurrence of these Kohn-Sham calculations within the MD simulation can be found in Fig.~\ref{fig:DftCumSteps}.  We observe that majority of the DFT calculations occur early during the MD, becoming less frequent as the simulation continues. Though similar observations have been made for ambient conditions \cite{jinnouchi2019fly, jinnouchi2020fly, kumar2023kohn},  a significantly larger number of DFT steps are performed even after a thousand MD steps here, as in previous on-the-fly MLFF MD simulations for transport properties of WDM \cite{kumar2024fly}, likely due to the increased movement of atoms causing new configurations to be encountered later in the MD simulation.  Even with these limited numbers of DFT calculations, they comprise on average 74\% of the total CPU time over the different temperatures, with a maximum of 95\% for 200 kK and a minimum of 50\% for 20 kK. In terms of overall performance, the speedup provided by the on-the-fly MLFF scheme relative to Kohn-Sham DFT is up to two orders of magnitude in both CPU and wall time, with an average speedup by a factor of 62, the minimum and maximum speedups being factors of 44 and 108, respectively.  Note that the current MLFF implementation has not been developed to scale in cases where the number of processors exceeds the number of atoms. Given that the MLFF code can be further parallelized to effectively utilize an order of magnitude more processors, and that the SPARC electronic structure code in default operation can scale to many thousands of processors \cite{xu2021sparc, zhang2023version}, and the Gauss SQ method can scale to many tens of thousands of processors \cite{suryanarayana2018sqdft, gavini2023roadmap, bhattacharya2022accurate}, the speedups presented here will also be achieved in the wall time of such simulations.

\begin{table}[htbp]
    \begin{tabular}{|c|c|c|c|c|c|}
        \hline
        \multirow{2}{*}{T [K]} & \multicolumn{2}{c|}{Time [CPU s]} & \multicolumn{2}{c|}{\# MD steps} &  \multirow{2}{*}{Speedup} \\
        \cline{2-5}
              &  MLFF    & DFT &   MLFF      & DFT & \\
        \hline
        10,000 & 47 & 4$\times 10^3$ & 9936 & 64 & 44 \\
        20,000 & 49 & 5$\times 10^3$ & 9925 & 75 & 46 \\
        50,000 & 45 & 7$\times 10^3$ & 9897 & 103 & 48 \\
        100,000 & 42 & 1$\times 10^4$ & 9873 & 127 & 48 \\
        200,000 & 21 & 1$\times 10^5$ & 9862 & 138 & 57 \\
        500,000 & 19 &  4$\times 10^4$ & 9872 & 128  & 61 \\
        750,000 & 19 & 2$\times 10^4$  & 9883 & 117 & 63 \\
        1,000,000 & 17 & 1$\times 10^4$ & 9901  & 99 & 69 \\
        2,000,000 & 15 & 1$\times 10^4$ & 9907 & 93 & 74\\
        5,000,000 & 12 & 5$\times 10^3$ & 9940 & 60 & 68\\
        10,000,000 & 11 & 4$\times 10^3$ & 9954 & 46 & 108\\
        \hline
    \end{tabular}
    \caption{Performance of the on-the-fly MLFF scheme for carbon at different temperatures. The timings correspond to each MD step, when averaged across the entire simulation. The speedup represents the ratio of the time taken for the entire MD simulation by Kohn-Sham DFT (extrapolated from 100 steps) and MLFF.}
    \label{Tab:MLFFPerformanceTable}
\end{table}

\subsection{\label{sec:FPEOScompare} Application: Shock Hugoniot for 14 materials}

We now calculate the shock Hugoniot for 14 materials from the FPEOS database, namely  9 elements: H, He, B, C, O, N, Ne, Mg, and Si; and 5 compounds: LiF, BN, B$_4$C, MgO, and MgSiO$_3$, at temperatures between 10 kK and 2 MK, containing both the WDM and HDM regimes. To facilitate comparison, the temperatures are chosen to be the same as those in the FPEOS database \cite{militzer2021firstprin}. The temperature for the initial state is chosen to be 300 K, and the densities for H, He, B, C, O, N, Ne, Mg, Si, LiF, BN, B$_4$C, MgO, and MgSiO$_3$ are chosen to be 0.085, 0.124, 2.465, 3.515, 0.667, 0.808, 1.507, 1.737, 2.329, 2.535, 2.258, 2.509, 3.570, and 3.208 g/cm$^3$, again as in the FPEOS database. In the MD simulations, we employ a time step of 0.63 fs for C at 10 kK, adjusting it for other temperatures and systems as the inverse square root of the temperature and square root of the mass of the lightest element in the system, respectively. We consider system sizes of 500 and 64 atoms for the lowest and highest temperatures of 10 kK and 2 MK, respectively, while interpolating linearly for temperatures in between. Due to stoichiometric constraints in the case of B$_4$C and MgSiO$_3$, the smallest system size is taken as 70 atoms.

\begin{figure*}[htbp]
\includegraphics[width=0.80\linewidth]{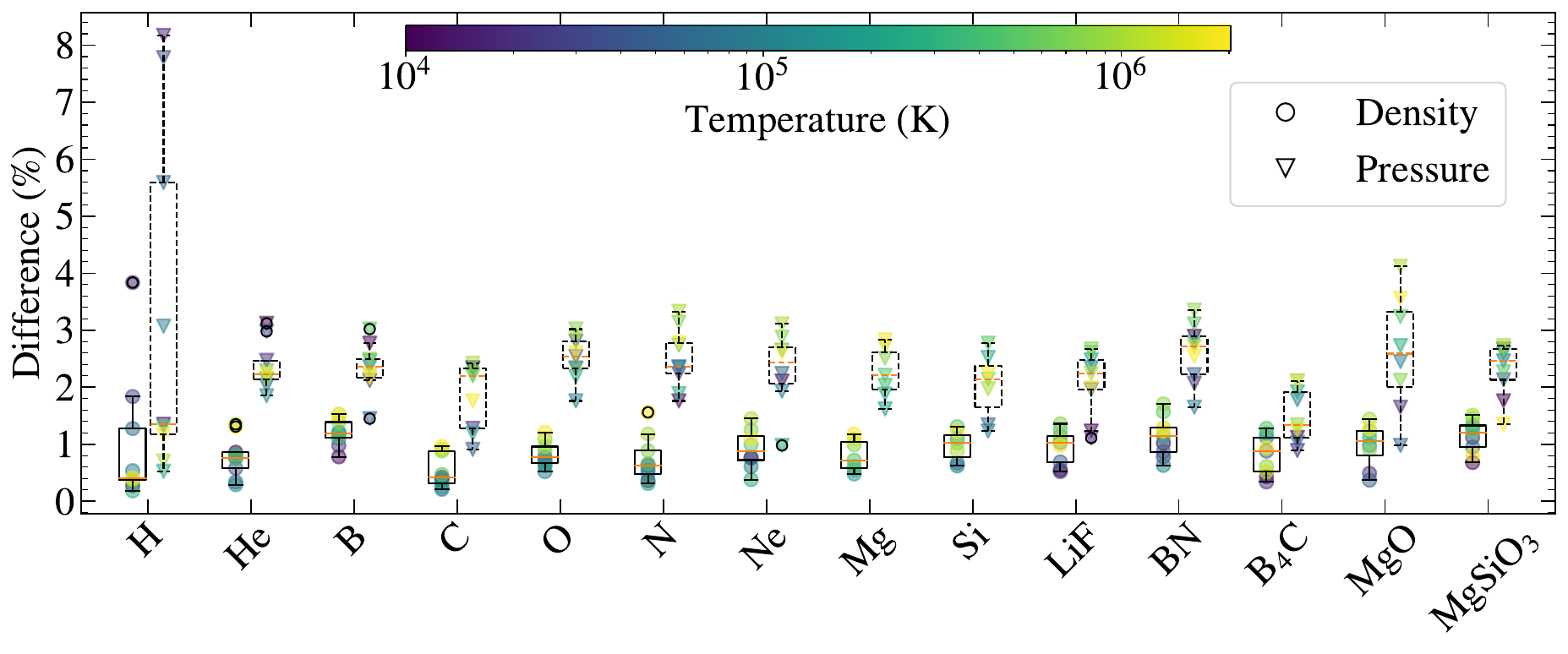}
        \caption{Difference in the density and pressure along the shock Hugoniots obtained from the on-the-fly MLFF scheme and the FPEOS database.}
        \label{fig:hug_error_FPEOS}
\end{figure*}

\begin{figure*}[htbp]
\subfloat[H]{\includegraphics[keepaspectratio=true,width=0.38\textwidth]{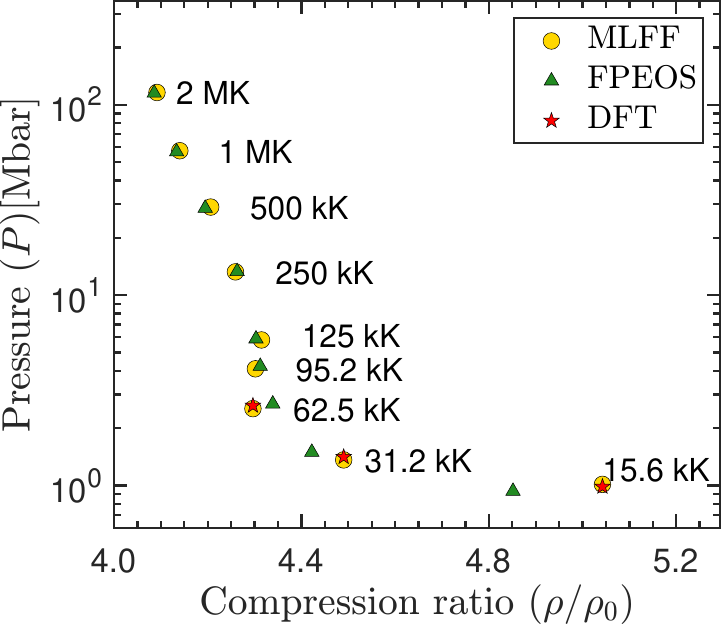} \label{Fig:hug_HC1} }
\hspace{0.04\textwidth}%
\subfloat[C]{\includegraphics[keepaspectratio=true,width=0.38\textwidth]{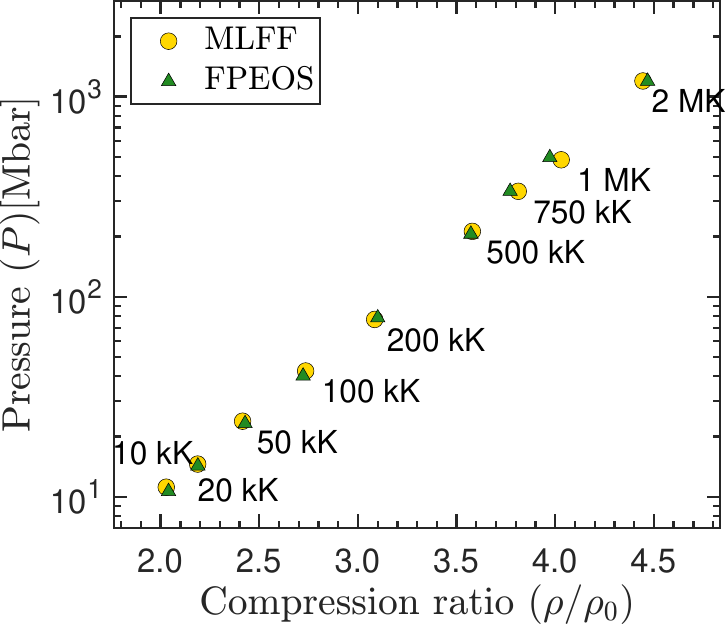} \label{Fig:hug_HC2} }
\caption{\label{Fig:hug_HC} Comparison of the shock Hugoniots obtained from the on-the-fly MLFF scheme and the FPEOS database for carbon and hydrogen.}
\end{figure*}

In Fig.~\ref{fig:hug_error_FPEOS}, we present the shock Hugoniots for the 14 materials computed by the on-the fly MLFF scheme and compare it with those obtained from the FPEOS database. The FPEOS database has Kohn-Sham DFT data for the lower temperatures and PIMC data for the higher temperatures, generally beyond 1 MK. The script provided within the database is used to generate the FPEOS Hugoniots. We observe that there is very good agreement for both the elements as well as the compounds, with the average difference in the density and pressure across all materials being 0.8\% and 2.5\%, respectively. The smallest and largest differences between the Hugoniots occur for C and H, respectively, which have been plotted in Fig.~\ref{Fig:hug_HC}. The larger differences in H, which occur at the lower temperatures, may be due to the fact that the FPEOS data at the lower temperatures are from PIMC simulations \cite{militzer2000pathPIMC, militzer2001path}, which is a many-body method with different approximations from those adopted in Kohn-Sham DFT. In particular,   PIMC calculations become more challenging at the lower temperatures and therefore may be  associated with larger error bars.  We also observe that the larger differences for other materials generally occur at the higher temperatures around 1 MK, which may again be attributed to the fact that the FPEOS data at such temperatures are from PIMC simulations. To verify that the relatively large differences for H are not a consequence of the MLFF model, we have performed Kohn-Sham DFT simulations at the three lowest temperatures of 15.625, 31.25, and 62.5 kK, with corresponding densities along the Hugoniot, i.e., 0.423, 0.369, and 0.365 g/cm$^3$, respectively. The results so obtained are also presented in Fig.~\ref{Fig:hug_HC1}. The errors in the pressure along the Hugoniot at these points are determined to be 0.62\%, 0.73\%, and 0.71\%, respectively, which further verifies the accuracy of the on-the-fly MLFF scheme. To check the sensitivity of the results to the choice of exchange-correlation functional, we also perform on-the-fly MLFF MD simulations  with LDA, and find that the results agree with those from PBE to within 0.52\%, 0.41\%, and 0.29\%, respectively.  To verify the accuracy of the chosen pseudopotential, we also perform on-the-fly MLFF MD simulations with the stringent ONCV pseudopotential from the PseudoDOJO library \cite{van2018pseudodojo}, and find that the results agree with those from the SPMS library to within 0.48\%, 0.37\%, and 0.46\%, respectively. Therefore, we find that the DFT results are insensitive to both choice of pseudopotential and choice of exchange-correlation functional; and hence that the relatively large differences from PIMC results for H at the lower temperatures are more likely due to the level of theory (semi-local exchange-correlation), larger error bars in the PIMC calculations, and/or other approximations such as nuclear-quantum effects.

\begin{figure}[htbp]
\includegraphics[width=0.90\linewidth]{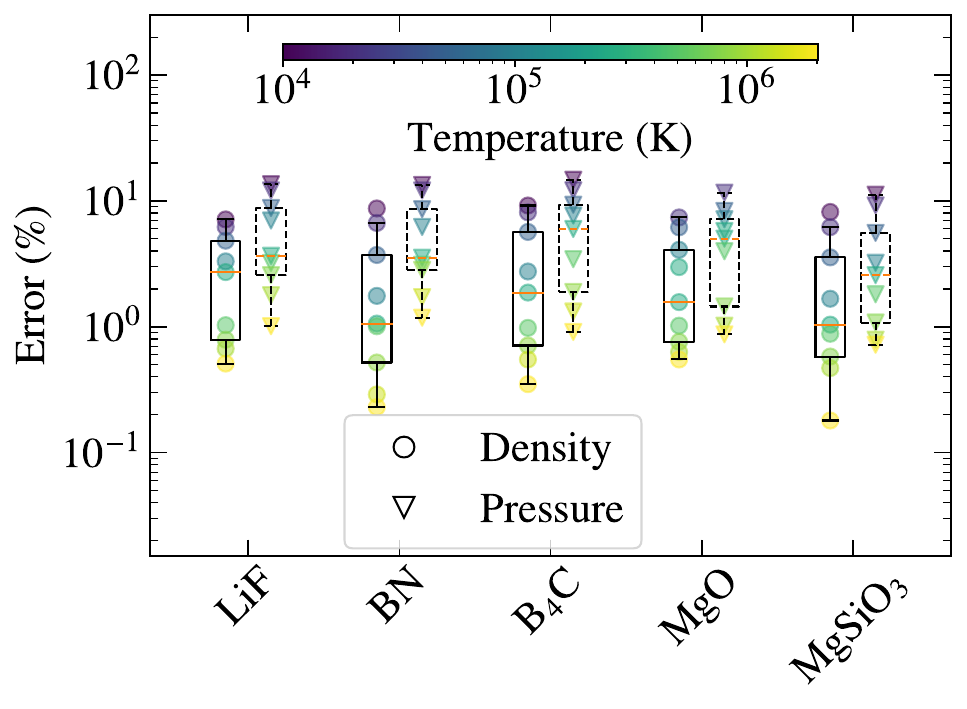}
        \caption{Error in the density and pressure along the shock Hugoniots within the linear mixing approximation for the compounds.}
        \label{fig:Error_ideal_mixing}
\end{figure}

In Fig.~\ref{fig:Error_ideal_mixing}, we present the error in the shock Hugoniots for the compounds, i.e.,  LiF, BN, B$_4$C, MgO, and MgSiO$_3$, when calculated using the linear mixing approximation. In the linear mixing approximation, the internal energy of a compound at a given temperature and pressure is approximated to be the sum of the internal energies of the individual elements, while the  density is approximated to be the weighted (by mass fraction) harmonic mean of the individual components. In so doing, we also compute the Hugoniots for Li and F using the on-the-fly MLFF scheme, since these have not been calculated as part of the comparison with FPEOS above. We observe that the linear mixing approximation becomes more accurate with increasing temperature, the errors being as large as 15\%  for the lower temperatures while  becoming smaller than 1\% for temperatures beyond 200 kK, which is consistent with previous results in the literature \cite{militzer2020nonideal}. This suggests that there is a reduction in inter-element interactions with temperature for the compounds studied here, essentially becoming negligible in the HDM regime.

\section{\label{sec:conclusion}Concluding remarks}
In this work, we have developed  a framework for computing the shock Hugoniot using on-the-fly MLFF with ab initio accuracy. In particular, we have employed an MLFF model based on the kernel method and Bayesian linear regression to compute the electronic free energy, atomic forces, and pressure,  alongside  an internal energy model based on linear regression of the electronic internal and free energies, both trained with Kohn-Sham DFT data.  We have verified the accuracy of the formalism by comparing the Hugoniot of carbon in the WDM and HDM regimes with recent Kohn–Sham DFT results in the literature.  In so doing, we have demonstrated that Kohn-Sham DFT calculations of the Hugoniot can be accelerated by up to two orders of magnitude, while retaining \emph{ab initio} accuracy.  We have applied this framework to calculate the Hugoniots for 14 materials from the FPEOS database, comprising 9 single elements and 5 compounds, over a temperature range from 10 kK to 2 MK. We have found that the results are in good agreement with the first principles results in the database while providing tighter error bars. In addition, we have verified that the accuracy of the linear mixing approximation for compounds improves as the temperature increases. 

The inclusion of temperature and density within the machine learned models will help further accelerate shock Hugoniot calculations, making it a worthy subject for future research. Extending the current MLFF implementation to further increase the scalability on large-scale supercomputers, along with GPU-accelerated computation of essential kernels \cite{sharma2023gpu}, promises to decrease wall time significantly in MD simulations, making it a promising area for future research.


\begin{acknowledgments}
S.K. and P.S. gratefully acknowledge support from grant DE-NA0004128 funded by the U.S. Department of Energy (DOE), National Nuclear Security Administration (NNSA). J.E.P gratefully acknowledges support from the U.S. DOE, NNSA: Advanced Simulation and Computing (ASC) Program at Lawrence Livermore National Laboratory (LLNL). This work was performed in part under the auspices of the U.S. DOE by LLNL under Contract DE-AC52-07NA27344. This research was also supported by the supercomputing infrastructure provided by Partnership for an Advanced Computing Environment (PACE) through its Hive (U.S. National Science Foundation through grant MRI-1828187) and Phoenix clusters at Georgia Institute of Technology, Atlanta, Georgia.
\end{acknowledgments}

\section*{Data Availability Statement}
The data that supports the findings of this study are available within the article, Supplementary Material, and from the corresponding author upon reasonable request.

\section*{Author declarations}
The authors have no conflicts to disclose.


\appendix

\begin{table*}[htbp]
    \centering

    \begin{tabular}{|c | c | c |c |c |c |c |c |c |c |c|}
        \hline
        \hline

        \hline
            \multirow{3}{*}{H} & T [kK]      &15.625 &31.25 &62.5 &95.25 &125 &250 &500 &1000 &2000 \\
          & $\rho$ [g/cm$^3$] &0.423 &0.369 &0.365 &0.367 &0.367 &0.362 &0.357 &0.352 &0.349  \\
          & P [Mbar]      &1.01 &1.36 &2.53 &4.12 &5.84 &13.24 &28.99 &56.21 &115.91  \\
         \hline

          \multirow{3}{*}{He} & T [kK]      &10 &20 &40 &60 &125 &250 &500 &1000 &2000 \\
          & $\rho$ [g/cm$^3$] & 0.403 & 0.434 & 0.542 & 0.588 & 0.654 & 0.642 & 0.587 & 0.558 & 0.531 \\
          & P [Mbar]      & 0.13 & 0.24 & 0.68 & 1.10 & 2.97 & 7.55 & 16.05 & 33.70 & 65.81 \\

          \hline
          
          \multirow{3}{*}{B} & T [kK]      &10 &20 &50.523 &101 &202.02 &505.05 &842 &1340 &2020 \\
          & $\rho$ [g/cm$^3$] &5.13 & 5.41 & 6.37 & 7.25 & 8.22 & 9.48 & 10.36 & 10.94 & 11.05 \\
          & P [Mbar]      &5.60 & 7.53 & 14.90 & 27.99 & 55.42 & 141.21 & 272.13 & 498.20 & 852.33 \\

          \hline

          \multirow{3}{*}{C} & T [kK]      &10 &20 &50 &100 &200 &500 &750 &1000 &2000 \\
          & $\rho$ [g/cm$^3$] & 7.20 & 7.69 & 8.51 & 9.58 & 10.85 & 12.56 & 13.41 & 14.21 & 15.62  \\
          & P [Mbar]      & 11.02 & 14.56 & 23.40 & 40.71 & 77.63 & 209.01 & 332.70 & 484.50 & 1192.95 \\

          \hline

          \multirow{2}{*}{C} & T [kK]      &10 &20 &50 &100 &200 &500 &750 &1000 &2000 \\
          \multirow{2}{*}{(LDA)}& $\rho$ [g/cm$^3$] & 7.21 & 7.68 & 8.51 & 9.58 & 10.85 & 12.56 & 13.40 & 14.20 & 15.62  \\
          & P [Mbar]      & 11.04 & 14.62 & 23.48 & 40.67 & 77.65 & 209.44 & 333.21 & 483.87 & 1194.26 \\

          \hline






          \multirow{3}{*}{O} & T [kK]    & 40 &50 & 80 &100 &250 &500 &750 &1000 &2020 \\
          & $\rho$ [g/cm$^3$] & 2.42 & 2.50 & 2.72 & 2.86 & 3.18 & 3.13 & 3.12 & 3.03 & 3.39 \\
          & P [Mbar]      & 1.29 & 1.62 & 3.06 &  4.32 & 16.20 & 38.70 & 66.39 & 88.07 & 245.14 \\

          \hline

          \multirow{3}{*}{N} & T [kK]      &10 & 20 &50 &100 &250 &500 &750 &1000 &2020  \\
          & $\rho$ [g/cm$^3$] & 2.88 & 3.03 & 3.05 & 3.17 & 3.46 & 3.50 & 3.53 & 3.54 & 3.98 \\
          & P [Mbar]      & 0.66 & 1.11 & 2.56 & 2.60 & 19.79 & 46.43 & 80.38 & 110.59 & 297.04 \\

          \hline
            
          \multirow{3}{*}{Ne} & T [kK]      &10 & 50 & 80 &100 &250 &500 &750 &1000 &2020  \\
          & $\rho$ [g/cm$^3$] & 3.15 & 4.55 & 5.24 & 5.47 & 6.77 & 7.11 & 7.22 & 6.86 & 6.85  \\
          & P [Mbar]      & 0.63 & 3.41 & 6.31 &  8.13 & 31.16 & 77.66 & 139.78 & 185.58 & 430.69 \\

          \hline

          \multirow{3}{*}{Mg} & T [kK]      &200 &250 & 300 &400 & 500 & 600 & 750 &1340 &2020 \\
          & $\rho$ [g/cm$^3$] & 6.97 & 7.41 & 7.75 & 8.14   &8.23 & 8.43 & 8.59 & 8.34 & 8.63 \\
          & P [Mbar]      & 22.02 & 30.69 & 40.11 & 58.21 & 77.74 & 102.42 & 144.24 & 305.64 & 538.65 \\

          \hline

          \multirow{3}{*}{Si} & T [kK]      &50 & 75 &100 & 200 &250 & 505.05 & 750 &1010 &2020 \\
          & $\rho$ [g/cm$^3$] & 5.96 & 6.65 & 6.94 & 8.04 & 8.68 & 10.61 & 11.30 & 11.55 & 11.43  \\
          & P [Mbar]      & 6.02 & 9.41 & 12.31 & 26.55 & 35.72 & 99.75 & 176.72 & 262.20 & 649.33 \\
          
        \hline
          
          \multirow{3}{*}{LiF} & T [kK]      &10 & 20 & 50  &100 &250 & 500 & 750 &1010 &2020  \\
          & $\rho$ [g/cm$^3$] & 5.36 & 5.90 & 6.98 & 8.06 & 9.85 & 10.99 & 11.08 & 11.42 & 11.73  \\
          & P [Mbar]      & 3.85 & 6.08 & 13.18 & 24.38 & 69.59 & 161.64 & 247.14 & 363.97 & 817.03  \\

          \hline
    
          \multirow{3}{*}{BN} & T [kK]      &10 & 20 & 50.50  &100 &252 & 505.05 & 842 &1340 &2020   \\
          & $\rho$ [g/cm$^3$] & 3.14 & 4.30 & 5.52 & 6.51 & 7.88 & 8.48 & 9.27 & 9.86 & 10.25  \\
          & P [Mbar]      & 1.02 & 3.36 & 9.43 & 10.24 & 60.55 &  124.65 & 240.68 & 433.53 & 766.79 \\

          \hline

          \multirow{3}{*}{B$_4$C} & T [kK]      &10 & 20 & 50.50  &100 &252 & 505.05 & 842 &1340 &2020  \\
          & $\rho$ [g/cm$^3$] & 4.95 & 5.41 & 6.34 & 7.28 & 8.66 & 9.32 & 10.60 & 11.15 & 11.61 \\
          & P [Mbar]      & 4.85 & 7.16 & 14.29 & 27.37 & 71.67 & 144.12 & 286.20 & 516.34 & 884.71 \\

            \hline

          \multirow{3}{*}{MgO} & T [kK]      &20 & 50 & 80 &100 &250 & 500 & 750 &1010 &2020 \\
          & $\rho$ [g/cm$^3$] & 7.62 & 8.88 & 9.68 &  10.06 & 12.57 & 14.29 & 14.69 & 14.91 & 16.27   \\
          & P [Mbar]      & 8.25 & 15.62 & 22.33 &  27.38 & 74.25 & 178.45 & 276.81 & 395.68 & 1073.14  \\

          \hline

          \multirow{3}{*}{MgSiO$_3$} & T [kK]      &10 & 20 & 50  &100 &250 & 500 & 750 &1010 &2020  \\
          & $\rho$ [g/cm$^3$] & 6.47 & 6.81 & 7.93 & 9.33 & 11.46 & 12.72 & 13.29 & 13.79 & 14.57  \\
          & P [Mbar]     & 4.41 & 5.95 & 12.47 & 24.26 & 65.48 & 155.93 & 256.06 & 384.86 & 902.67  \\
 
        \hline    
        \hline 
    \end{tabular}
    \caption{Hugoniot data for the materials studied in this work, as determined using the on-the-fly MLFF scheme.}
     \label{Tab:Data}
\end{table*}

\section{\label{App:Data} Hugoniot data}
The Hugoniot data for the material systems studied in this work is summarized in Table \ref{Tab:Data}. The complete EOS data can be found in the Supplementary material.

%


\end{document}